# Development of Practical Soft X-ray Spectrometers.

G. Brammertz, P. Verhoeve, A. Peacock, D. Martin, N. Rando, R. den Hartog and D. J. Goldie.

*Abstract*—Cryogenic soft X-ray imaging spectrometers are currently being developed for applications in the fields of astronomy and material sciences. In this paper we present experiments on optimized single devices, which show measured energy resolutions of 4.6 eV, 8.1 eV and 20.5 eV at 525 eV, 1.5 keV and 6 keV respectively. These energy resolutions combined with a quantum efficiency of more than 40 % in the energy range from 0.5 to 2 keV together with a count rate capability of 15 kHz demonstrate the overall good performance of single Superconducting Tunnel Junctions (STJs). Assembling these optimized single devices in a matrix read-out would provide the practical basis for a soft X-ray imaging spectrometer.

## I. INTRODUCTION

FUTURE X-ray missions such as XEUS [1] intend to use cryogenic imaging spectrometers in their focal plane. For XEUS the good inherent energy resolution and quantum efficiency of these detectors will be exploited to acquire spectra of even the most distant and youngest known objects in the universe so as to determine their redshift, age and composition.

Another possible application for these spectrometers is in the field of material analysis, where high energy resolutions are required for X-ray diagnostics, that provides spatially resolved microchemical analysis [2].

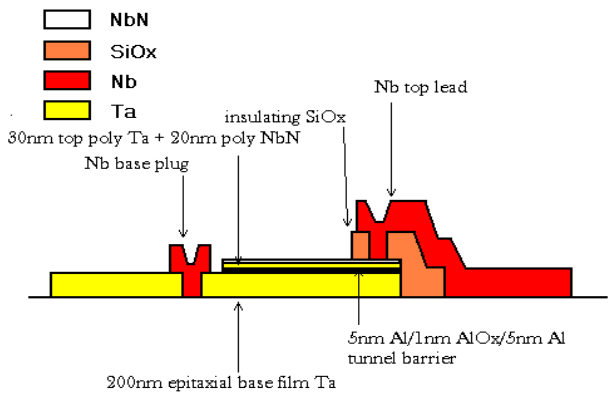

Fig. 1. General layout of the Superconducting Tunnel Junction described in this paper.

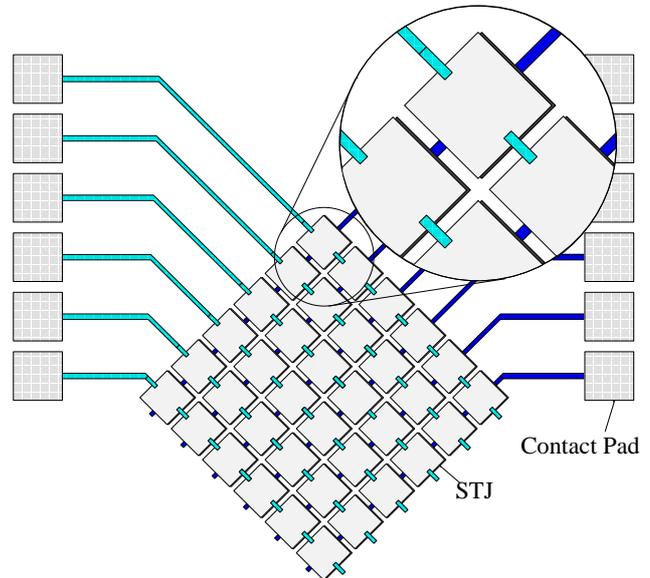

Fig. 2. Principle of the matrix read-out scheme: All top films of a line and all base films of a column are connected to a common bias and processing circuit [7].

In both cases high energy resolution is needed to resolve overlapping or complex line structures.

Currently silicon based energy-dispersive spectrometers are limited by an energy resolution of about 100 eV at 1 keV. While wavelength dispersive spectrometers provide a good energy resolution, they are severely limited by their overall efficiency.

Cryogenic imaging spectrometers based on Superconducting Tunnel Junctions (STJs) do not suffer from any of these limitations. Their good intrinsic resolution and quantum efficiency at soft X-ray energies (<2keV) make them ideal detectors for these kind of applications.

In this paper we will provide the characteristics of single STJs, which have been optimized for the soft X-ray region, and we will show that the assembly of these single devices into matrix arrays is feasible.

## II. SUPERCONDUCTING TUNNEL JUNCTIONS

### A. Principle of operation

A STJ consists of two superconducting films separated by an insulating layer (see Fig. 1). A magnetic field is applied parallel to the barrier to suppress the Josephson current. The absorption of a photon of energy E in one of these two superconducting electrodes is followed by a series of fast processes in which the photon energy is converted first into high energy phonons, which then create a number $N_0$ of quasi-particles by Cooper pair breaking. The mean number of quasi-particles created is $N_0=E/1.7\Delta$, where $\Delta$ is the energy gap of the superconductor. These quasi-particles can be read-out by applying a d.c. bias voltage across the insulating barrier. The quasi-particles tunnel across the barrier, causing a charge transfer from one electrode to the other, for which the direction is determined by the polarity of the applied bias voltage. After a first tunnel through the barrier the quasi-particles can back-tunnel, causing another charge to be added to the signal. In this way every quasi-particle contributes on average $<n>$ times to the signal before it is lost in the system. The number of electrons collected in the signal current therefore is $N=<n>N_0$. In theory, the best achievable energy resolution of symmetrical STJs is the so-called tunnel-limited resolution given by [3], [4]:

$$\delta E_{tun} = 2.355\left[1.7\Delta E(F+1+\frac{1}{<n>})\right]^{1/2}, \qquad (1)$$

where F is the Fano factor [5], [6]. In practice the energy resolution is further degraded by additional components such as spatial non-uniformities in the detectors response ($\delta E$ proportional to E) and electronic noise ($\delta E$ independent of energy).

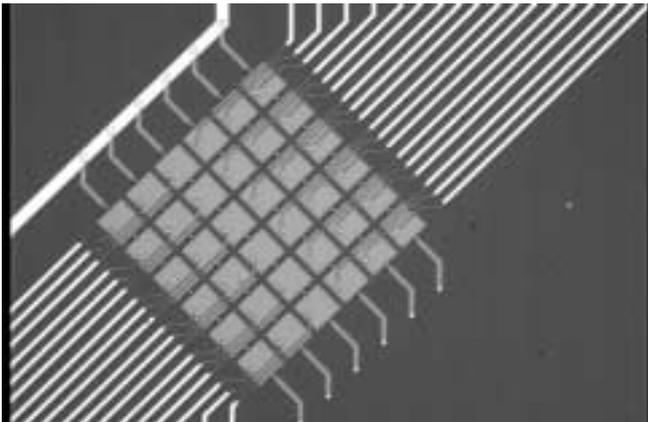

Fig. 3. Single device read-out scheme for a 6x6 array. Note the numerous top contact leads ($N^2$) obscuring the pixels. Each base lead is however connected to a common ground and each devices Ta base film is interconnected electrically to its neighbor via a narrow bridge plugged with Nb. The Nb plug allows electrical connectivity but reduces quasi-particle diffusion out of the device due to the higher energy gap material.

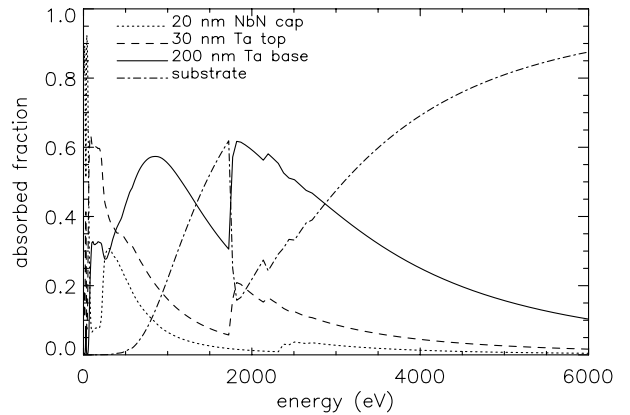

Fig. 4. Absorbed fraction of incident photons in the different films of the STJ for the device layout presented in this paper.

*B. Matrix read-out*

The single pixel read-out of STJs requires an electrical connection to the base and to the top electrode of every junction. This causes obstruction problems of the STJs by the numerous leads when large arrays of STJs need to be read-out. Also $N^2$ signal processing chains are needed for an NxN array. An alternative approach is the so-called matrix read-out scheme, where all the base films of a column and all the top films of a row are connected together [7] (see Fig. 2). Every row and every column is connected to its own bias and amplification circuit. When a pixel is hit by a photon a signal will be produced in its corresponding row and column, allowing the exact determination of the events position. This read-out scheme eliminates the obscuration of the top film by the top lead and reduces the number of processing chains to 2N. Drawbacks however are the lower count-rate capabilities and larger electronic noise due to the higher capacitances of the N junctions connected in parallel.

### III. OPTIMIZING DETECTION EFFICIENCY

Generally the best energy resolution is obtained in the high quality epitaxial base film of our STJs. In order to prevent photons from being absorbed in the polycrystalline top film before reaching the base film, the thickness of this top film, associated leads and the dielectric Silicon Oxide layer used to insulate the top film contact from the base film should be kept as thin as possible. Here we describe how the detection efficiency can be optimized in a large STJ array.

In a conventional STJ read-out scheme, where every junction is read out separately, the top leads and the insulating Silicon oxide layers obstruct large parts of the area of the STJ (see Fig. 3). This obstruction increases with the increasing number of pixels to be read-out. For the matrix read-out the top leads are only necessary from one edge of a STJ to the edge of the adjacent STJ (see Fig. 2). The obstruction of the device by the leads is thus very small. Also the insulating Silicon oxide must only be deposited over part of the edges of the device, underneath the top lead, and not over the complete array, as in the case of large arrays with single pixel read-out. The most recent experimental results on this new read-out scheme are reported in [8].

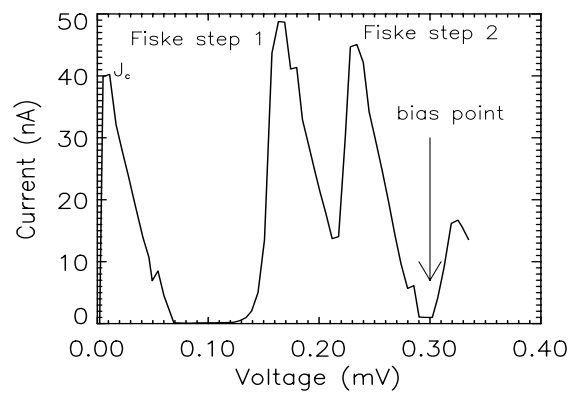

Fig. 5. IV-curve of the 30 µm x30µm device under examination. Note the low leakage currents (~ 0.22 pA/µm$^2$) and the low voltages of the first Fiske steps. $J_c$ represents the Josephson current.

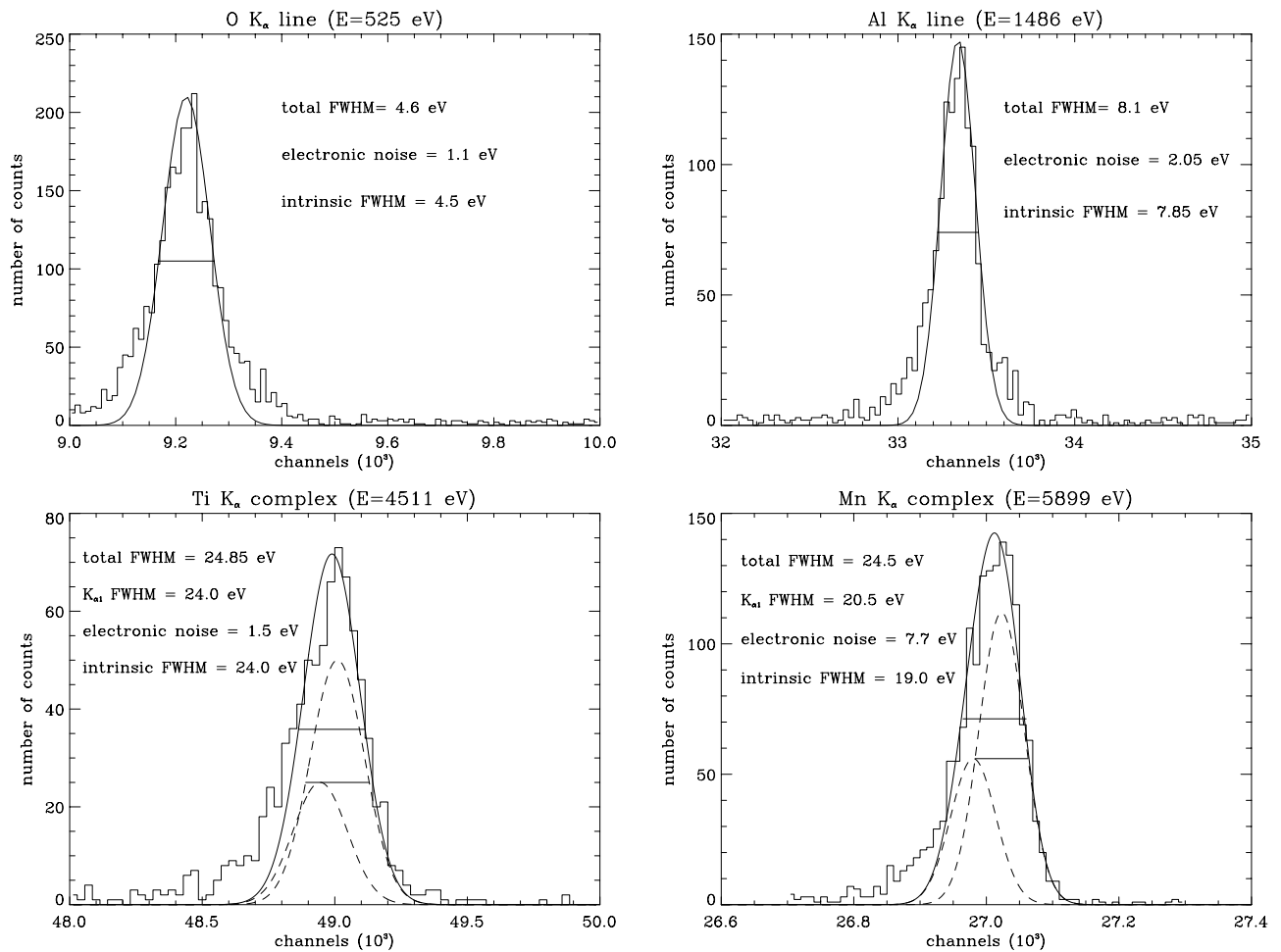

Fig. 6. Spectra for the O, Al, Ti and Mn $K_\alpha$ lines. Fits to the obtained spectra are also indicated. For the Ti and Mn lines the fits to the data were made by including both the $K_{\alpha 1}$ and $K_{\alpha 2}$ lines of the $K_\alpha$ complex.

In this paper we show the results from a single Ta based STJ with a thick base and a thin top film, optimized for maximum photon absorption in the base film. The lay-up of the device is: 200 nm Ta/5 nm Al/1 nm AlOx/5 nm Al/30 nm Ta/20 nm NbN. The high energy gap NbN film was deposited in order to prevent the quasi-particles from reaching the free surface on which an oxide layer naturally forms, which may represent an important quasi-particle loss source [9],[10]. However from comparison with similar devices without these NbN layer, no evidence for a reduction of these losses was found. Also the measured energy resolution for devices with and without the NbN layer was not found to be significantly different. We conclude therefore that the NbN layer is not essential for the present devices. Fig. 4 shows the detection efficiency for the top and base films of this optimized lay-up as a function of incident photon energy. Note that the detection efficiency in the base film is 40 to 50 % over the 0.5 to 2 keV energy range. The thin NbN layer only has a significant effect below 0.5 keV. A further increase in base film thickness to 500nm, which is quite feasible, would improve the detection efficiency at 2 keV to ~ 80 %. Improved efficiency below 1 keV would require however further thinning of the top film, which is not straightforward (see section IV).

## IV. Experimental Results

The experiments presented in this paper were performed in a portable $^3$He cryostat with a base temperature of ~ 360 mK. The read-out electronics consist of a charge sensitive pre-amplifier at room temperature, followed by a two-channel shaping stage, determining rise-time and charge output for every pulse. The photons were generated with a Manson X-ray source, having a choice between 6 different targets: Al, Mn, C, LiF, Cu and Ti. The results at 5.9 keV were obtained with a radioactive $^{55}$Fe sample in a different He$^3$ cryostat with a base temperature of 300 mK. The device tested is a single

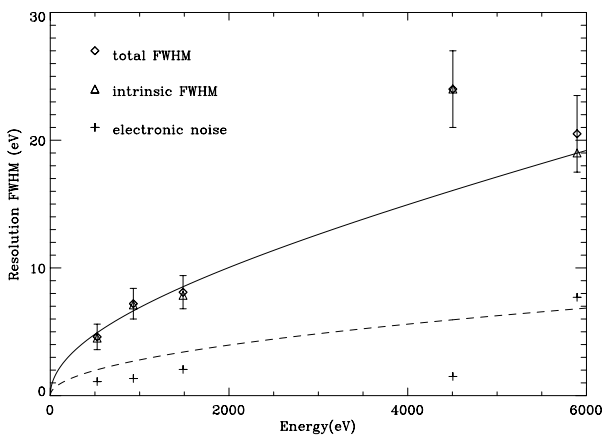

Fig. 7. Energy resolution as a function of incident photon energy. The dashed line represents the tunnel limited energy resolution. The solid line represents a least mean squares fit to the intrinsic FWHM resolution using an equation of the form $\sqrt{aE + bE^2}$ with a=0.0448 eV$^{-1}$ and b= 2.77 10$^{-6}$ eV$^{-2}$.

30 μm x 30 μm STJ, which shows a leakage current of 0.22 pA/μm$^2$ at 100 μV and a normal resistance of 1.5 10$^{-6}$ Ωcm$^2$. The mean energy gap, as determined from the IV-curve, of the two superconducting multilayers is 0.68 meV. A magnetic field of ~ 150 Gauss was applied parallel to the junction to suppress the Josephson current.

Fig. 5 shows part of an I-V curve of the junction. The first Fiske step for this device appears at a bias voltage as low as 160 μV. This is due to the thin top film bilayer, with a thickness of only 50nm. This makes stable biasing of the device more difficult, especially for larger junctions, where the Fiske steps are at even lower bias voltages. For a comparable 30x30 μm$^2$ device with an equal base and top film thickness of 100 nm the first Fiske step appears only at a voltage of 300 μV. It is for this reason that the top films cannot be made arbitrarily thin.

Spectra from the different X-ray sources were acquired. Fig. 6 shows the O (which occurs as a contamination line in the spectra from all targets), Al, Ti and Mn K$_\alpha$ lines, as detected with the junction. The intrinsic energy resolutions (after subtraction of the electronic noise) are: 4.5 eV for O-K$_\alpha$ (525 eV), 7.85 eV for Al-K$_\alpha$ (1486 eV), 24 eV for Ti-K$_{\alpha1}$ (4511) and 19 eV for the Mn K$_{\alpha1}$ line (5899 eV). These energy resolutions have been corrected for a non-linearity in the energy response of the detector. They still include, however, the natural widths of the X-ray emission lines, which are of the order of 1-3 eV, and possibly an additional broadening due to the source, in the case of O-K$_\alpha$, Al-K$_\alpha$ and Ti-K$_\alpha$. Fig. 7 shows the total measured energy resolution as a function of incident photon energy. Also indicated are the electronic noise levels (as determined from the electronic pulser width) in each measurement and the intrinsic energy resolution, obtained by subtracting in quadrature the electronic noise from the measured energy resolution. The dashed line represents the tunnel limited resolution, whereas the solid line is a least mean squares fit to the intrinsic energy resolution. The fit was made using an equation of the form $\sqrt{aE + bE^2}$. The parameters a and b deduced were respectively 0.0448 eV$^{-1}$ and 2.77 10$^{-6}$ eV$^{-2}$. Generally, the intrinsic resolution is about a factor 2-3 above the predicted tunnel limit. Note, the results at high energies appear relatively better than at low energy, which may be indicative of some residual broadening of the source spectra at low energies due to the X-ray source. This residual broadening may also be the reason for the relatively high a parameter obtained from the fit to the data. From equation (1) one would expect a 5 to 6 times lower value for a.

V. CONCLUSIONS

We have presented photon counting experiments with a single Superconducting Tunnel Junction optimized for the soft X-ray region (<2 keV). The layout of the present junctions, having a 200 nm thick base film and a 50 nm top film, is designed to favor the detection efficiency in the high quality epitaxial base film of the detector. A further optimization of the detection efficiency is possible by increasing the base film thickness. The measured intrinsic energy resolutions in the base film corrected for electronic noise, but not for natural line width and line broadening due to the X-ray source, are 4.5 eV for the O (525 eV), 7.85 eV for the Al K$_\alpha$ line (1486 eV) and 19 eV for the Mn K$_{\alpha1}$ (5899 eV) line. This type of device, when combined with a matrix read-out scheme would provide the basis for an imaging soft X-ray spectrometer. Problems with photon absorption in the overlying films are minimized, as a result of the thin top film and, for a matrix array, the absence of Silicon oxide and top leads over the major part of the spectrometer.